\documentclass[10pt, conference, letterpaper]{IEEEtran}

\usepackage{enumitem}

\usepackage{cite}
\usepackage{graphicx}
\usepackage{caption}
\usepackage{subcaption}
\usepackage{refstyle}
\usepackage[cmex10]{amsmath}

\usepackage{algorithmic}
\usepackage{array}
\usepackage{url}
\usepackage{amsbsy}
\usepackage{fontenc}
\usepackage[utf8]{inputenc}
\usepackage{mathtools}
\usepackage{tabularx}
\usepackage{tabu}
\usepackage{float}
\usepackage{gensymb}
\usepackage{multirow}
\usepackage{siunitx}
\usepackage{bbold}
\usepackage{algorithmic}
\usepackage{psfrag}
\usepackage{pgfplots}
\usepackage{xcolor}

\usepackage{comment}
\usepackage{wrapfig}
\usepackage[utf8]{inputenc}
\usepackage{amsmath,amssymb}
\usepackage{makecell}
\usepackage{booktabs}
\usepackage{bm}
\usepackage{enumitem}
\usepackage{dblfloatfix}    
\usepackage[ruled,vlined]{algorithm2e}
\usepackage{mathtools,nccmath}

\SetKwInput{KwInput}{Input}                
\SetKwInput{KwOutput}{Output}              
\usepackage{flushend}

\newcommand{\navid}[1]{{\color{teal}{\bf #1}}}

\newsavebox{\ieeealgbox}

\newcolumntype{M}[1]{>{\arraybackslash}m{#1}}
\newcolumntype{N}{@{}m{0pt}@{}}

\usepackage{proof}

\begin{document}


\title{
Digital-Twin assisted Network Energy Optimization during Low Traffic Hours}

\author{
\IEEEauthorblockN{Shuvam Chakraborty, Ahmed Bedewy, Wenjun Li, Navid Abedini}
\IEEEauthorblockA{
$^*$Qualcomm Technologies Inc. - e-mail: \{shuvchak, ahmebede, wenjunl, navida\}@qti.qualcomm.com}
\vspace{0mm}

}

\maketitle
\pagestyle{empty}

\begin{abstract}

As wireless network technology advances towards the sixth generation (6G), increasing network energy consumption has become a critical concern due to the growing demand for diverse services, radio deployments at various frequencies, larger bandwidths, and more antennas. Network operators must manage energy usage not only to reduce operational cost and improve revenue but also to minimize environmental impact by reducing the carbon footprint. The 3rd Generation Partnership Project (3GPP) has introduced several network energy savings (NES) features. However, the implementation details and system-level aspects of these features have not been thoroughly investigated.
In this paper, we explore system-level resource optimization for network energy savings in low-traffic scenarios. We introduce multiple NES optimization formulations and strategies, and further analyze their performance using a detailed network digital twin. Our results demonstrate promising NES gains of up to 44\%. Additionally, we provide practical considerations for implementing the proposed schemes and examine their impacts on user equipment (UE) operation.
\end{abstract}


\section{Introduction}

The global proliferation and deployment of 5G New Radio (NR) have remarkably met the ever-increasing demand from users and applications with unprecedented data rates~\cite{5G_survey}. To achieve quality of service (QoS) requirements such as reliability, low latency, and enhanced capacity, the total cost of ownership, including capital expenditure (CAPEX) and operational expenditure (OPEX), has significantly increased in modern wireless networks.
According to recent reports by the Global System for Mobile Communications Association (GSMA)~\cite{gsma_report} and the Next Generation Mobile Networks (NGMN) Alliance~\cite{9349624}, on average, 90\% of the energy use for a telecommunications operator — one of the main contributors to OPEX — comes from the network side, with the radio access network (RAN) accounting for 50-80\% of the total network energy use. Therefore, optimizing the energy efficiency of the RAN is paramount for the telecommunications industry as we move forward to 6G. 

{Statistical analysis on typical traffic distribution across the network, such as the one presented in~\cite{nokia_report}, reveals a very non-uniform traffic load distribution across time and base stations.}
It is observed that around 70\% of base stations (gNB) handle only 20\% of the traffic, resulting in very low utilization. Additionally, traffic during peak hours is approximately 60-70\% higher than the average and significantly greater than during low-traffic hours at night. {Consequently, many base stations, during few hours per day, are in no or low-traffic state. During this period, the base stations are still active and consume energy, mostly for providing system broadcasts (non-data traffic) and maintaining idle resources.}
This provides a significant opportunity to design and implement NES features targeting low-traffic and idle scenarios.

3GPP is currently exploring idle-mode NES strategies for standardization in 5G Advanced and 6G. However, the practical implications and system-level implementation aspects have not been thoroughly examined. Optimizing network operations for idle-mode requires information about the idle UEs that are not connected to the network. Traditional methods allow the network to track the idle UEs only at the granularity of tracking areas, which typically encompass multiple cells. To enable efficient network optimization at the cell or sub-cell levels, advanced technologies such as digital twins (DT)~\cite{DT} are needed. { 
In the context of wireless communication networks, a DT is a high fidelity digital representation of the network that provides an accurate representation of the radio frequency and physical environments using computer vision based methods~\cite{jiang2024learnablewirelessdigitaltwins} and ray tracing techniques~\cite{ray_tracing}, as well as the network nodes' behaviors and operations.}

This work presents a comprehensive system-level analysis aimed at minimizing the OPEX related to energy consumption during idle mode. The optimization focuses on network resources, specifically the set of active cells and the associated communication beams.
The key contributions of this work are as follows:
\begin{enumerate}[leftmargin=*, nosep, topsep=0pt]
    \item We propose system-level NES strategies in off-peak hours by selecting optimal configurations for cells and beams.

    \item We develop a simulation framework based on a highly accurate digital twin of an urban network deployment.
        
    \item Using the developed DT, we analyze the proposed strategies in terms of their impact on the network OPEX and demonstrate NES gains of up to 46.4\% in the mentioned network.
    
    \item Finally, we provide practical considerations for implementing the proposed NES strategies and analyze their impact on UE operation.
\end{enumerate}

The remainder of the paper is organized as follows. \S\ref{sec:related} reviews previous work related to this paradigm. \S\ref{sec:prelim} introduces essential topics and context necessary for the core system design. 
\S\ref{sec:system} details the  optimization problem formulations.
\S\ref{sec:sim} presents the developed digital twin framework. 
~\S\ref{sec:results} evaluates the proposed strategies. 
Finally, \S\ref{sec:conclusion} concludes the paper. 

\section{Relevant Works}
\label{sec:related}

Exploring the existing discussions on NES and energy efficiency (EE), the authors in~\cite{10121451} investigate various techniques for reducing network power consumption by adapting time, frequency, spatial, and power domain resources. They also provide an analysis of the trade-offs between EE and throughput to identify which domain offers the highest potential for power savings.
An extensive EE study on multiple-antenna cellular networks is conducted in~\cite{8454484}, introducing a more robust concept of area EE instead of spatial EE.
An overview of general industry practices for RAN energy saving, including both 3GPP and open RAN, is provided in~\cite{kundu2024energyefficientranindustry}, highlighting the trend towards AI/ML-based designs in 6G.
{A deep reinforcement learning based solution for NES has been proposed in~\cite{low_traffic_nes_samsung}, where base stations develop local energy saving solutions.}
The authors in~\cite{8922617} survey new and anticipated architectural changes in 6G networks aimed at network sustainability, ubiquitous coverage, pervasive AI/ML application, and enhanced protocols.
An analytical and system-level study of various densification technologies for 6G networks is performed in~\cite{azzino2024energycostefficient6gnetworks}, focusing on EE analysis for traffic-loaded scenarios.
The authors in~\cite{deployement_paper} showcase a DT-assisted network planning and deployment optimization problem, minimizing network CAPEX, which is unique among existing works. 
{None of the previous studies explore the DT-assisted network wide energy optimization for low-traffic and idle scenarios, and at different levels of granularity.}


\section{Preliminaries}
\label{sec:prelim}
\subsection{SSB Codebook} \label{sec: SSB prelim}
In 5G, signal transmission and reception can be directed towards specific beam directions. Broadcast control channels, such as the Synchronization Signal Block (SSB), System Information Blocks (SIB), and Physical Random Access Channel (PRACH), may be supported at multiple instances with different beam directions to ensure coverage of entire cell area.

The set of beams associated with broadcast control channels is sometimes referred to as the SSB codebook. According to 3GPP specifications, it is possible to configure up to 8 SSBs in frequency ranges below 6 GHz and up to 64 SSBs in the millimeter-wave frequency range. However, the design of the SSB codebook is determined by the network implementation, which decides the number of SSB beams (within the specified limits), as well as their azimuth and elevation spans and beam directions.
Figure~\ref{fig:baseline_codebook} illustrates example of an SSB codebook comprising 32 beams that uniformly divide an angular range of 120 degrees in azimuth and 30 degrees in elevation.
\vspace{-10pt}
\subsection{Energy Consumption Model}
In this work, the energy consumption model adopted for gNB is based on the 3GPP specifications \cite{3gpp.38.864}. The model includes various inactive and active modes, featuring three sleep levels (micro, light, and deep) for the inactive mode, as well as active downlink (DL) and active uplink (UL) states. For the active states, the power expenditure (P) of the gNB consists of two distinct parts:
\smallskip
\begin{equation}\label{eq:power_model}
    P = P_{static} + P_{dynamic} 
\end{equation}
The static component ($P_{static}$) is present regardless of the gNB's configuration, and is assumed to have the same value as the micro-sleep power consumption. The dynamic component ($P_{dynamic}$) 
 depends on the gNB's configuration, such as antenna configuration, transmit power, and occupied bandwidth. In reality, a gNB can comprise multiple cells. However, for simplicity in this work, we assume each gNB has one cell and hence may use the term cell and gNB interchangeably. 
 
Based on this model, we calculate a cell's energy cost in idle mode, where it transmits SSBs and SIBs, and monitors for PRACH with a given periodicity (e.g., 20 msec).
For a given cell, with $N_b$ number of active beams, the energy cost can be represented as:
\begin{equation}\label{eq:cell_opex}
    C(N_b) = 1_{\{N_b>0\}}c_{\textrm{static}} + f(N_b)
\end{equation}
where, $1_{\{X\}} = 1 ~\textrm{if}~ X$ is TRUE and $0$ otherwise; $c_{\textrm{static}}$ is a constant energy cost per cell and $f(x)$ is the additional energy cost of supporting $x \ge 1$ beams in the SSB codebook.

\begin{figure}
\centering
\includegraphics[width=0.8\linewidth]{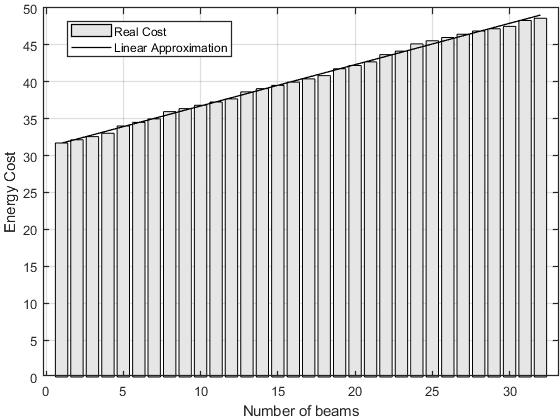}
\caption{Cost of operation for a cell in idle mode as a function of number of SSB beams.}
    \label{fig:sample_ssb_period}
    \vspace{-20pt}
\end{figure}

Figure~\ref{fig:sample_ssb_period} shows energy cost $C(N_b)$ for various number of beams $N_b$, and assuming 20 msec periodicity of the control channels. 
More information about the other assumptions for this analysis is provided in~\S\ref{sec:sim}.
From this figure, we can approximate $f(N_b)$ as a linear function. Hence:
\begin{equation}\label{eq:cell_opex_approx}
    C(N_b) \approx 1_{\{N_b>0\}}c_{\textrm{static}} + mN_b
\end{equation}
where $m$ is the approximate cost of each active beam.

\section{System-level Network Energy Optimization}
\label{sec:system}

\subsection{Baseline Network Deployment}
An initial deployment of the cell sites is selected to meet a \textit{target throughput} for a peak UE load.
Specifically, we assume that in the geographical area under consideration, there are potential active UE locations (associated with all possible outdoor locations) 
and candidate cell sites (associated with available poles in the area). 
The goal of the initial network deployment problem is to choose the optimal cell sites 
that satisfy the \textit{target throughput} requirement during peak load time, while minimizing the total deployment cost. 
Each cell site is assumed to comprise three sectors (or cells) that divide the site's coverage region. 
This problem is solved based on an a priori signal-to-interference-plus-noise ratio (SINR) estimation, which dictates the datarate a serving cell can provide to a user. 

{For the distribution of the users, we assume $N_{TP}$ potential traffic points (user locations) are distributed uniformly with a resolution of 1 square meter area for each in the outdoor areas and $N_p$ number of pole locations are identified as potential cell sites. To meet a target throughput of $r_t$ 
for each user and assuming a cell can serve $K$ users (termed multiplexing factor) with equal resource (duty cycle) scheduling, each user must have an achievable channel rate of $r_tK$. Based on the target throughput, an SINR threshold SINR$_{th}(K)$ can be obtained as a function of $K$.  
The link SINR between each cell and traffic point is estimated using ray-tracing as well as an estimated interference margin. 
For a given candidate cell and multiplexing factor $K$, the number of surrounding traffic points with link SINR to the cell exceeding SINR$_{th}(K)$ is denoted by $P(K)$. 
The activity factor $a$ is defined as the percentage of traffic points with an active UE. The number of traffic points that can be supported with activity factor $a$ and multiplexing factor $K$ is given by $Q(K) = K/a$. 
Therefore, the optimal multiplexing factor that yields the highest number of supported traffic points for a cell is given by $K^{*} =  \arg\max_{k} (\min(P(K), Q(K)))$, from which we obtain the selected SINR threshold SINR$_{th}(K^{*})$. 
Based on selected SINR thresholds for all candidate cell locations, a binary connectivity matrix $\bm{A}_{dep}$ of dimension $(N_{TP},N_p)$ is defined, 
where $A_{ij} = 1$ if candidate cell $j$ and TP $i$ has a link exceeding the SINR threshold estimated for that cell.}
The general optimization problem is formulated as minimizing the number of deployed cells such that a given percentage $\alpha$ of traffic points (under the assumed activity factor) achieve the target throughput: 
\begin{align}\label{eq:deployment}
   \min_{\bm{x}_p} {\bm{1}^T_{N_p}}{\bm{x}_p} , ~s.t. ~ \bm{A}_{dep}\bm{x}_p \ge \bm{y}, ~ {\bm{1}^T_{N_{TP}}}\bm{y} \ge \alpha
\end{align}
{where $\bm{y}$ is a binary vector of length $N_{TP}$ with $y_i$ indicating TP $i$ is covered and $\alpha$ is a predetermined \% target throughput coverage for the network deployment. 
We encourage the reader to review ~\cite{deployement_paper} for a more in-depth discussion on the optimization method adopted to solve this deployment problem.}

While the selection of deployed cell sites 
aims to satisfy \textit{capacity} requirements during the peak times, the network should primarily ensure a minimum \textit{coverage} for UEs during off-peak hours. 
The \textit{coverage} requirement is defined in terms of the necessary link budget for broadcast control channels, such as SSB, SIB, PRACH, which is typically less stringent than the SINR requirements for data traffic (associated with \textit{capacity} needs).
Therefore, although the initially selected cell sites can provide coverage to $N_{UE}$ users (or more accurately, $N_{UE}$ user locations in the area), fewer network resources (e.g., cells, or beams) may be sufficient to maintain the same \textit{coverage} during off-peak hours. This presents an opportunity for network energy savings by activating only subset of cells and/or beams during these times.

A UE $j$ is considered to be covered by a cell $i$ (or a beam $i$), if the associated SINR meets an SSB SINR threshold: $\textrm{SINR}_{ij} \ge \textrm{SINR}_{th}$. 
Accordingly, we construct a connectivity matrix ($\bm{A}$) to represent the coverage: 
\begin{align}\label{eq:connectivity}
    A_{ij} &= 1~~\textrm{if}~ \textrm{SINR}_{ij}>\textrm{SINR}_{th} \\
             &= 0~~\textrm{otherwise.} \nonumber 
\end{align}
The general network energy saving problem in idle mode can be formulated as follows,
\begin{equation}\label{eq:high-level}
    \min_{\bm{x}} cost(\bm{x}), ~s.t. ~ {\bm{A}^T}\bm{x} \ge \bm{1}_{N_{UE}}
\end{equation}
in which, $\bm{x}$ is a binary vector whose $i^{th}$ element indicates whether a cell $i$ (or beam $i$) is active, $\bm{1}_{N}$ is an all-one vector of length $N$, and $cost(\bm{x})$ is the energy cost associated with the decision vector $\bm{x}$.

The optimization problem (\ref{eq:high-level}) can be formulated and solved at different levels of granularity: 1) local beam (de)activation (i.e., selecting a set of active beams for each cell independently), 2) global cell (de)activation (i.e., selecting a set of active cells), and 3) global cell and beam (de)activation (i.e., selecting a set of active cells and the active beams within them). Next, we elaborate on each of these optimization strategies, which are also outlined in Table~\ref{tab:nes_strat}.  

\begin{table}[]
\begin{center}
\vspace{10pt}
\begin{tabular}{|c|c|}
\cline{1-2} 
\textbf{Strategy}        & \textbf{Description}      \\ \cline{1-2} 
  \multicolumn{2}{|c|}{\textbf{Local Solution}}                \\ \cline{1-2}  
Beam Level          & Optimizing SSB codebook per cell \\
\cline{1-2} 
  \multicolumn{2}{|c|}{\textbf{Global Solution}}                \\ \cline{1-2}  
Cell level &   Optimizing the set of active  
  cells, \\
  & each cell using the baseline SSB codebook \\
\cline{1-2}
 Beam and cell level   &   Jointly optimizing the set of active cells    \\
&  and SSB codebook for each active cell  \\
\cline{1-2}
\end{tabular}
\vspace{5pt}
\caption{NES strategies in idle mode.}
\label{tab:nes_strat}
\end{center}
\vspace{-25pt}
\end{table}

\subsection{Local Beam-level Optimization}
As discussed in Section~\ref{sec: SSB prelim}, each cell may adopt a baseline SSB codebook for the transmission and reception of its common channels. 
By leveraging an accurate DT, we can determine the radio coverage region of each cell, and identify the potential location and signal strength of the UEs associated with them, based on a UE-cell association criterion such as the strongest received power or SINR.
This analysis reveals that the coverage region of cells can be highly non-uniform, meaning that that minimum SINR among the associated UEs in different beam directions may vary. For example, there may be blockage in certain directions, preventing UEs from being associated beyond a certain distance, or above a path-loss threshold. Non-uniform coverage can also result from signal strength-based UE-cell association in overlapping coverage areas of neighboring cells. Figure~\ref{fig:min_snr} illustrates such a spatially non-uniform coverage area of a cell from our DT. 

SSB codebook for each cell $c$ can be optimally selected to minimize energy cost while meeting the cell's target coverage region. We assume each cell $c$ has a pool of $N_{B,c}$ beams, including beams of different shapes (and hence different gains) and directions to serve the entire coverage area of its associated $N_{UE,c}$ user locations.
 Given that the energy cost in (\ref{eq:cell_opex}) is an increasing function of the number of beams, minimizing the number of beams for a given cell is sufficient to minimize the energy cost of the associated cell. Considering a beam-level connectivity matrix $\bm{A}_{beam, c}$ of dimension ($N_{B,c}, N_{UE,c}$) for a given cell $c$, the local beam optimization problem is formulated as:
\begin{equation}\label{eq:beam_per_sector}
    \min_{\bm{x}_{b}} \bm{1}^T_{N_{B,c}}\bm{x}_b, ~s.t. ~ {\bm{A}^T_{beam,c}}\bm{x}_b \ge \bm{1}_{N_{UE,c}}
\end{equation}
wherein $\bm{x}_b$ is the binary decision vector, of length $N_{B,c}$, indicating the set of active beams. 
It is important to note that in this variation of the optimization problem, we assume the UE-cell association information is given (e.g., determined based on the DT and highest SINR based cell selection criterion). This means that the same set of UE locations will be covered by a cell, after the beam optimization. 

\subsection{Global Optimization}
Optimizing network energy consumption globally for the idle mode can be performed at two different levels of granularity: cell-level and joint cell- and beam-level. One of the main differences between global and local optimization problems is that, for global optimization, we do not assume any prior UE-cell association; instead, all possible connections are considered as part of the optimization.

\subsubsection{Cell-level}
Let $N_C$ indicate the number of originally deployed and active cells. The global cell-level optimization identifies a subset of these cells that can collectively cover all $N_{UE}$ user locations with minimum total energy cost.

For simplicity, we assume each cell uses a baseline SSB codebook comprising $N_{B,base}$ beams. However, the problem formulation can be straightforwardly extended to support different numbers of baseline beams for different cells. According to~\eqref{eq:cell_opex}, the energy cost of each active cell is $C(N_{B,base})$. Therefore, minimizing the total energy cost is equivalent to minimizing the number of active cells. Using~\eqref{eq:connectivity}, we consider a connectivity matrix ($\bm{A}_{cell}$) of dimension $(N_C, N_{UE})$ between cells and UEs. The global cell-level optimization problem is formulated as:
\begin{equation}\label{eq:problem_sc}
    \min_{\bm{x}} \bm{1}_{N_C}^T\bm{x}, ~s.t. ~ {\bm{A}_{cell}^T}\bm{x} \ge \bm{1}_{N_{UE}}
\end{equation}
wherein  $\bm{x}$ is the binary decision vector, of length $N_C$, indicating the set of active cells.
 
\subsubsection{Joint Cell- and Beam-level} 
The problem formulations so far aim to minimize the total energy costs globally at cell-level and locally by optimizing the beam selection per cell. 
By combining both approaches, a complete solution can be provided by selecting an optimal set of beams (and associated cells) from a network-wide pool of beams, of size $N_C N_B$, that can cover all $N_{UE}$ locations while minimizing total energy cost.
Following~\eqref{eq:connectivity}, we consider a connectivity matrix ($\bm{A}_{beam}$) of dimension $(N_C N_B, N_{UE})$ between all beams and UEs. 
We further define a new binary matrix $\bm{B}$ of dimension ($N_C N_B, N_C$) to indicate the association between beams and cells. Specifically,  $B_{ij}=1$, if the $i^{th}$ beam in the network-wide pool of beams is associated with cell $j$, and zero otherwise.  
The binary vector $\bm{x}$, of length $N_C N_B$, is used to indicate the set of selected active beams. Given these parameters, the total number of active beams in the network is $\bm{1}_{N_CN_B}^T\bm{x}$, and the set of active cells (i.e., cells with at least one active beam) can be represented by the indicator vector $\bm{1}_{\{\bm{B}^T\bm{x}>0\}}$. The number of active cells is thus $\bm{1}_{N_C}^T\bm{1}_{\{\bm{B}^T\bm{x}>0\}}$. 

Using the approximated energy cost function in~\eqref{eq:cell_opex_approx}, the global joint optimization can be formulated as:
\begin{align}\label{eq:nonlin_prob}
    &  \min_{\bm{x}}~ c_{static}\bm{1}_{N_C}^T\bm{1}_{\{\bm{B}^T\bm{x}>0\}}+ m\bm{1}_{N_CN_B}^T\bm{x}, \\
    &s.t. ~ {\bm{A}_{beam}^T}\bm{x} \ge \bm{1}_{N_{UE}} \nonumber
\end{align}

This objective function is not a linear functions of $\bm{x}$ due to the indicator function in the first term of the energy cost.
Theoretically, we can solve this as a non-linear problem. However, given the scale of the problem in large, densified networks, finding an optimal solution may be impractical.

{To linearize the problem, we utilize a relaxation approach from partial set cover problems~\cite{chekuri2019approximatingpartialsetcover}. Typically, such relaxation is applied to linear or integer linear programming where an auxiliary variable is introduced to satisfy additional constraints for the solution space~\cite{Bragin2022}. We can apply the same concept here to replace} the non-linear indicator function $\bm{1}_{\{\bm{B}^T\bm{x}>0\}}$ with an auxiliary binary decision vector $\bm{x}_c$ that indicates which cells are active. We further add a new linear constraint relating $\bm{x}_c$ and $\bm{x}$ as follows. Let $N_b(i)\in[0,N_B]$ denote the number of active beams for cell $i$, i.e., $N_b(i) := \bm{B}^T\bm{x}$. If $N_b(i)=0$ (or $>0$), then we require $\bm{x}_c(i)=0$ (or $1$). These conditions will be met by the optimal $\bm{x}_c$ using the following constraint: $N_B\bm{x}_c \ge \bm{B}^T\bm{x}$. 
 The modified problem is formulated as:

\begin{align}\label{eq:main_problem}   
    &\min_{\bm{x},\bm{x}_c}~ c_{static}\bm{1}_{N_C}^T\bm{x_c} + m\bm{1}_{N_CN_B}^T\bm{x}, \\
    &s.t. ~ {\bm{A}_{beam}^T}\bm{x} \ge \bm{1}_{N_{UE}}, N_b\bm{x}_c \ge \bm{B}^T\bm{x} \nonumber
\end{align}
\section{Digital Twin for System-level Analysis}
\label{sec:sim}

\begin{figure}
\centering
\includegraphics[width=0.9\linewidth]{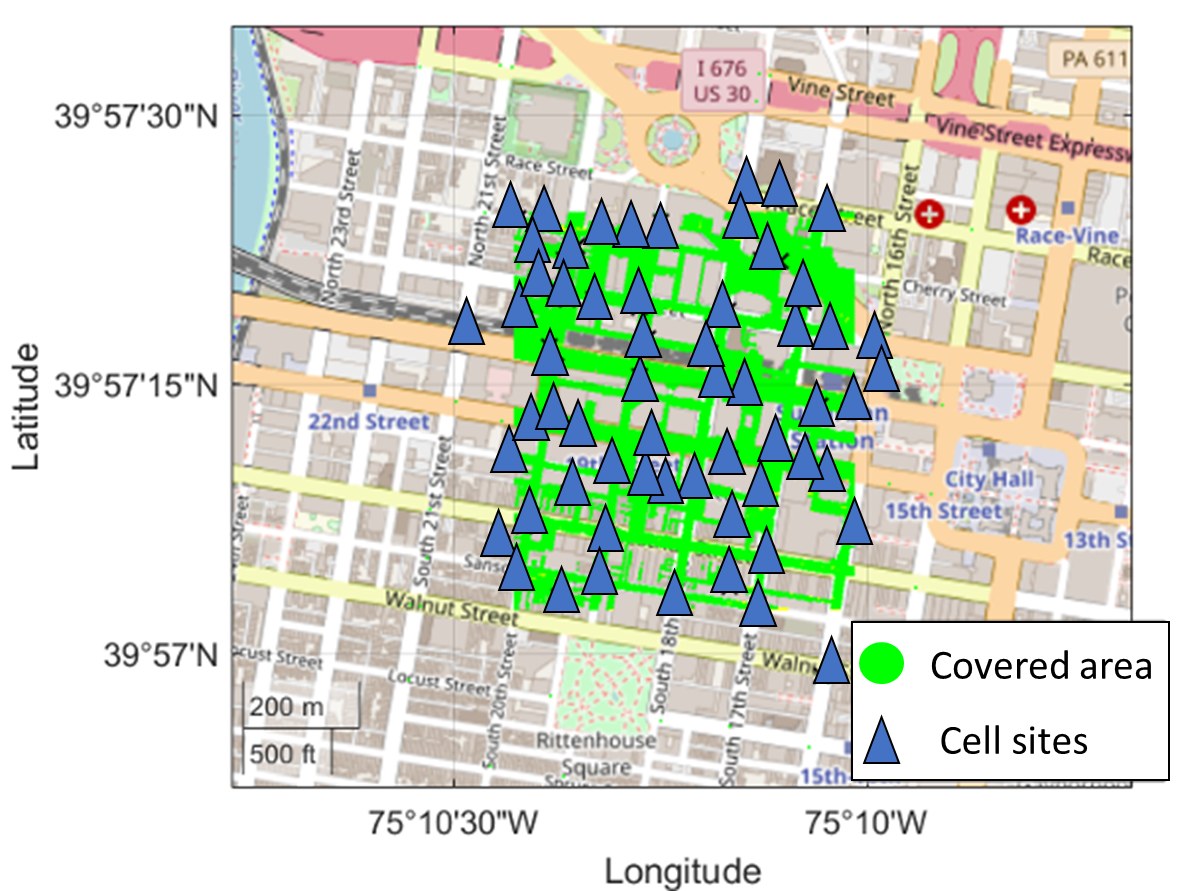}
\caption{28GHz network based on Downtown Philadelphia Digital Twin \small{(Map data \copyright  OpenStreetMap contributors, Microsoft, Esri community Maps contributors. Map layers by Esri, markers added for more visibility, license: https://creativecommons.org/licenses/by-sa/2.0/legalcode)}}
    \label{fig:network}
    \vspace{-15pt}
\end{figure}

We develop a detailed digital twin of an area (0.34 km $\times$ 0.28 km) in downtown Philadelphia (shown in Figure~\ref{fig:network}), accurately representing a cellular network operating at millimeter wave frequencies. 
To analyze the network's capacity and coverage, we model only outdoor UE locations, which are uniformly distributed across 1 m $\times$ 1 m grids in outdoor areas.
The radio frequency channels are generated based on ray tracing using parameters provided in Table~\ref{tab:rf_parameters}, along with realistic foliage modeling in the DT and the associated losses. 

The network under consideration consists of 49876 UE locations and $N_C=169$ active cells distributed across 69 sites, meeting the \emph{throughput} target of 50 Mbps for 80\% of UEs in the initial deployment. The same set of cells provides SSB \emph{coverage} to 94\% ($N_{UE}=46884$) of all UEs.

\begin{table}[]
\begin{center}
\vspace{10pt}
\begin{tabular}{|c|c|c|}
\cline{1-3} 
\textbf{Parameters}        & \textbf{Cell}         & \textbf{UE}      \\ \cline{1-3} 
{Carrier Frequency}        &  \multicolumn{2}{|c|}{28 GHz}                \\ \cline{1-3}    
{Bandwidth}        &  \multicolumn{2}{|c|}{800 MHz}                \\ \cline{1-3}  
{Antenna array } & 8 row x 24 col  & 2 row x 2 col  \\
{arrangement} & x 2 polarization & x 2 polarization \\ \cline{1-3}
{Maximum array gain} & 28.15 dBi & 10 dBi \\   \cline{1-3}  
{Body Loss} & N/A & 8 dB \\  \cline{1-3}  
{Implementation margin} & N/A  & 1.9 dB  \\ \cline{1-3}  
{Noise Figure}  & 10 dB & 6.7 dB  \\ \cline{1-3}  
{Cell-edge reliability margin}  & \multicolumn{2}{|c|}{13.2 dB}                \\ \cline{1-3}  
{Foliage loss} &  \multicolumn{2}{|c|}{4 dB/m}                \\ \cline{1-3}  
{Building reflection loss} & \multicolumn{2}{|c|}{6.4 dB}                \\ \cline{1-3} 
\end{tabular}
\caption{Digital-twin parameters}
\label{tab:rf_parameters}
\end{center}
\vspace{-25pt}
\end{table}

\begin{figure*}
\centering
\begin{subfigure}[b]{0.34\linewidth}
    \includegraphics[width=\linewidth]{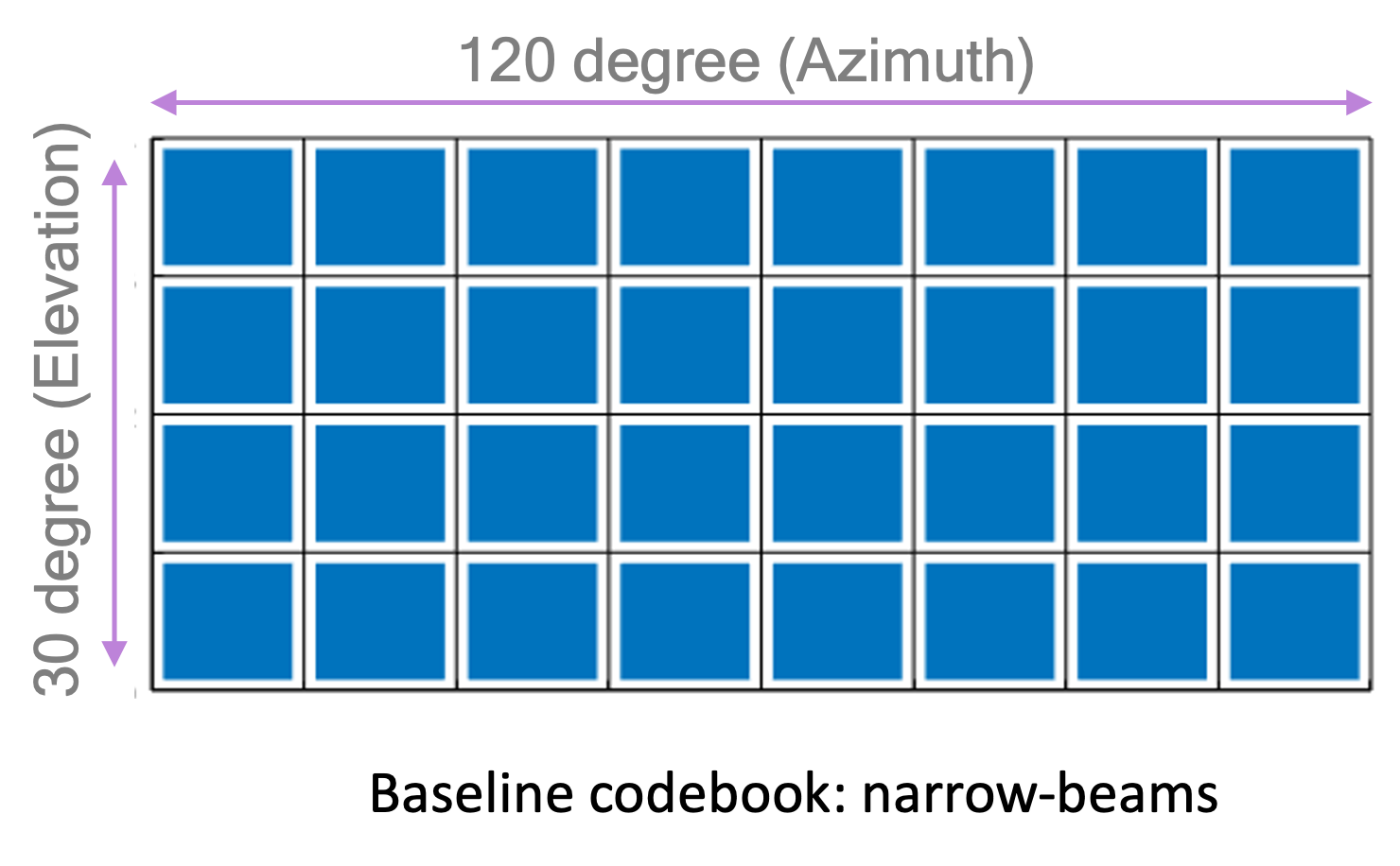}
    \caption{Baseline SSB codebook arrangement.}
    \label{fig:baseline_codebook}
\end{subfigure}
\hfill
\begin{subfigure}[b]{0.31\linewidth}
    \includegraphics[width=\linewidth]{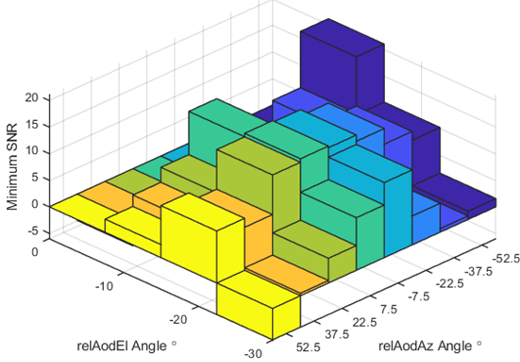}
    \caption{Minimum SNR along beam directions.}
    \label{fig:min_snr}
\end{subfigure}
\hfill
\begin{subfigure}[b]{0.31\linewidth}
   \includegraphics[width=\linewidth]{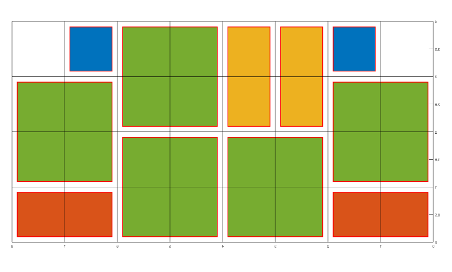}
\caption{Sample of an optimized SSB codebook.}
    \label{fig:sample_hybrid_beam}
\end{subfigure}
\caption{Baseline SSB codebook (a), and opportunities for SSB codebook optimization in (c) for a given cell with a non-uniform coverage (b).}
\label{fig:beam_per_sector}
\vspace{-15pt}
\end{figure*}

The baseline SSB codebook comprises 32 SSB beams, each with a width of 15$^{\circ}$ in azimuth and 7.5$^{\circ}$ in elevation, covering an azimuth span of   120$^{\circ}$ (-60$^{\circ}$ to 60$^{\circ}$) and an elevation span of 30$^{\circ}$ (-30$^{\circ}$ to 0$^{\circ}$), wherein 0$^{\circ}$ elevation is associated with the horizon, and negative values represent angles below the horizon. Figure~\ref{fig:baseline_codebook} shows the arrangement of the beams in the baseline codebook.

For the beam-level optimizations (both local and global), we consider four different types of beams with varying beamwidths in azimuth (BW$_{Az}$) and elevation (BW$_{El}$); Type-1: (BW$_{Az} =$ 15$^{\circ}$, BW$_{El} =$ 7.5$^{\circ}$), 
Type-2: (BW$_{Az} =$ 15$^{\circ}$, BW$_{El} =$ 15$^{\circ}$),
Type-3: (BW$_{Az} =$ 30$^{\circ}$, BW$_{El} =$ 7.5$^{\circ}$), and
Type-4: (BW$_{Az} =$ 30$^{\circ}$, BW$_{El} =$ 15$^{\circ}$).
Figure~\ref{fig:sample_hybrid_beam} shows an example of an SSB codebook comprising beams of different types, optimally selected for a cell with a non-uniform coverage region (as shown in Figure~\ref{fig:min_snr}).
The NES optimizations presented in~\S\ref{sec:system} are solved within the integer solution space using a mixed integer linear program (MILP)~\cite{10.1287/ijoc.2018.0857}.

\section{Evaluation}
\label{sec:results}
In this section, we present the performance results of the proposed system-level optimizations using the developed DT. We also discuss practical aspects of implementing these strategies, and evaluate their impacts on the UE operation.


\begin{table*}[]
\begin{center}
\vspace{10pt}
\begin{tabular}{|c|c|c|c|c|c|}
\cline{1-5} 
\textbf{Strategy}  & \textbf{Number of Active Cells} & \textbf{Number of Active Beams}  & \textbf{Energy Consumption (relative unit)} & \textbf{Energy Saving (\%)} \\ \cline{1-5} 
Baseline    &212           & 6784             &9.494$\times 10^3$   & N/A   \\ \cline{1-5}  
Local beam-level       &212          & 1002             &7.031$\times 10^3$   & 25.9   \\ 
optimization~\eqref{eq:beam_per_sector}         &            &           &         &   \\ \cline{1-5}
Global cell-level       &163          & 5216             &7.300$\times 10^3$   & 23.0   \\ 
optimization~\eqref{eq:problem_sc}         &            &           &         &   \\ \cline{1-5}
Global joint beam- and      &163          & 490             &5.286$\times 10^3$   & 44.0   \\ 
cell-level optimization~\eqref{eq:main_problem}      &            &           &         &  \\ \cline{1-5}
\end{tabular} 
\vspace{5pt}
\caption{Network energy savings with system-level optimizations at different level of granularity.}
\label{tab:results}
\end{center}
\vspace{-10pt}
\end{table*}
\vspace{-5pt}
\subsection{Network Energy Saving} 
A comprehensive set of results is presented in Table~\ref{tab:results}, indicating optimizations at different levels of granularity.
The table also references the equations corresponding to different optimization formulations.

We observe that locally minimizing the number of active beams for each cell can reduce total number of beams by 85.2\% 
leading to 25.9\% network energy saving.
A globally optimal cell ON-OFF strategy deactivates 23\% of cells, offering proportional network energy saving. 
However, a global strategy that optimally selects the set of active cells, and their associated active beams guarantees the most energy savings, up to 44\%. This global strategy suggests
shutting down 23\% of the cells and deactivating up to 92.8\% of the total SSB beams in the network.
As expected, increasing the level of granularity and solving the optimization problem globally offers significantly greater gains than coarse and/or local optimizations. Next, we will discuss the practicality of implementing such strategies. 


\subsection{Practicality of the Proposed Strategies}
Global optimization of system-level resources relies on a key assumption that an accurate DT is available, specially for the scenarios involving idle or inactive UEs where the network does not have sufficient knowledge about their location, population, and link qualities. A well-calibrated DT can be used to provide a reliable representation/expectation of such missing information.  
Depending on the use-case, and the operator's choice, the level of fidelity of a DT can be selected to balance the complexity of creating the DT and the achievable gains. 
For example, a simple radio environment map of the network with potential gNB locations and assuming uniform user distribution can allow a cell or even beam level optimization. However, If the network has access to a DT calibrated based on the historical user distribution, beam, scheduling, and other network behavior information, higher NES gains are possible.


Implementation of the proposed optimizations can be based on a central network management entity, such as a Service Management and Orchestration (SMO) system. This allows for the optimal selection of cells and beams for (de)activation, to achieve the highest NES. 
In practice, such an entity may produce a centralized action to be implemented, or a policy to be followed by gNBs. The policy can be expressed in terms of a set of time periods (within a day) and/or thresholds (e.g., traffic load or number of connected UEs), along with associated actions (e.g., cell/beam (de)activation) for each cell.

Another aspect related to the centralized optimization is the scale of the optimization problem. Although referred to as global optimization, in reality, the entire operator's network should be divided into clusters for performing such optimization. A cluster can range in size from a whole metropolitan area to a local neighborhood. The choice should balance the trade-off between complexity and the ultimate reward of the network optimization.

Alternatively, the optimizations can be performed in more localized areas and in a distributed manner with local information sharing across neighboring gNBs. Additionally, simpler heuristics for network optimizations may be pursued, where gNBs take iterative actions. The implementation and analysis of such scenarios are beyond the scope of this work and are left for future research.
\begin{figure*}
\centering
\begin{subfigure}[b]{0.325\linewidth}
    \includegraphics[width=\linewidth]{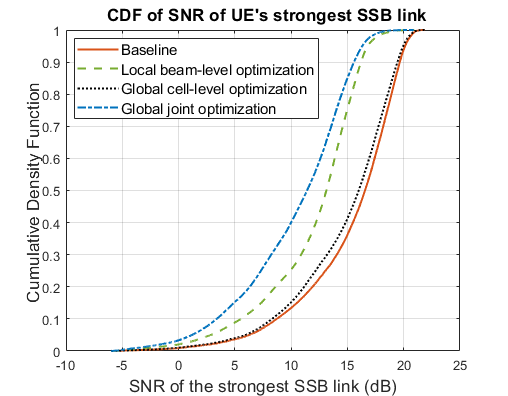}
    \caption{Distribution of SSB SNR.}
    \label{fig:dl_ssb_snr}
\end{subfigure}
\begin{subfigure}[b]{0.325\linewidth}
    \includegraphics[width=\linewidth]{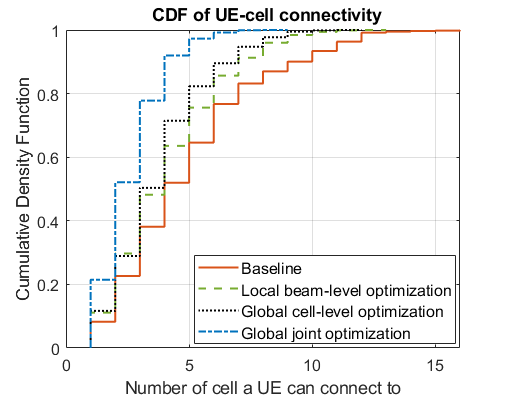}
    \caption{Coverage diversity.}
    \label{fig:coverage_div}
\end{subfigure}
\begin{subfigure}[b]{0.325\linewidth}
    \includegraphics[width=\linewidth]{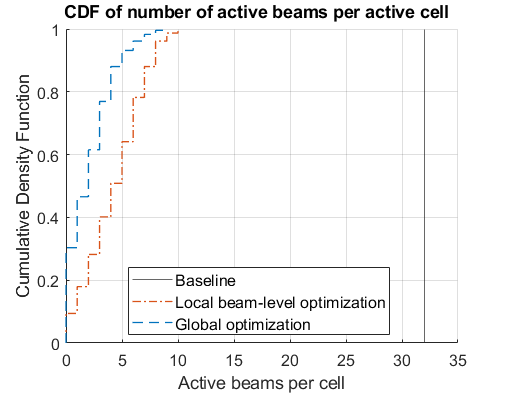}
    \caption{Number of active beams per cell.}
    \label{fig:beam_per_sector}
\end{subfigure}
\caption{Impact of the proposed NES strategies on UE operation.}
\label{fig:UE_imapct}
\vspace{-15pt}
\end{figure*}

\vspace{-5pt}

\subsection{On Complexity of the Solution}
It is important to note that the decisions for (de)activation of cells and beams at low or no traffic scenarios should essentially be made in a slow time scale (to avoid frequent changes in the system), and are based on historical or statistical assumptions about the possible idle UE locations and their link qualities. 
Hence, the nature of the problem is not dynamic (unlike problems like scheduling that depend on connected UEs' random and dynamic locations, mobility and traffic). As such, the proposed optimizations, in this work, are to be performed in non-real-time. This makes the concerns about associated complexities less relevant. However, for the sake of completeness a time complexity discussion of the proposed methods are provided below.
The optimization variables being in integer spaces, the problems are solved as MILP. Treating MILP as an optimization method, its complexity is NP-hard. However, in numerical solutions, if branch and bound method is applied, each branch can be considered to be solved in polynomial time as they are deterministic decision problems in each branch.
With 234 active cell in initial deployment along with 49876 potential UE locations in an area spanning 1.512 square kilometer, the time required to find optimal solutions, using the same machine, for the three NES solutions are provided in Table \ref{tab:results}. We refrain from reporting time complexity for the local beam-level optimization, as the scale of the problem in local level and network-wide optimization is different.

\begin{table}[]
\begin{center}
\vspace{10pt}
\begin{tabular}{|c|c|c|c|c|c|}
\cline{1-2} 
\textbf{Strategy} &\textbf{Computation time}\\ 
 &\textbf{(seconds)}\\ \cline{1-2} 
Baseline   & N/A  \\ \cline{1-2}  
Global cell-level        & 2.53 \\ 
optimization~(7)          & \\ \cline{1-2}
Global joint beam- and     & 34.25 \\ 
cell-level optimization~(9)       &\\ \cline{1-2}
\end{tabular} 
\vspace{5pt}
\caption{Computation time for different NES solutions with the network under consideration.}
\label{tab:results}
\end{center}
\vspace{-10pt}
\end{table}
\vspace{-5pt}

\section{Impact on UE Operation}
Although the optimization problems are formulated with a coverage constraint to ensure that no coverage hole are created as a results of cell/beam deactivations, they may still impact the operation of idle UEs, as elaborated further below.

\subsubsection{Link SNR}
Deactivating some cells/beams forces the associated UEs to be served by neighboring cells/beams with lower signal quality and strength. This may degrade the operation of the idle UEs in practice, as they may need multiple attempts to successfully receive a DL signal (such as SSB, SIB, or the paging message), or transmit an UL signal (such as PRACH). 
Figure~\ref{fig:dl_ssb_snr} shows the reduction in the SSB SNR of the strongest link associated with each UE after different local/global optimizations. 

\subsubsection{Coverage Diversity} 
To maintain reliability, coverage diversity is a desirable feature where a given UE can be covered by more than one cells and/or beams. However, the proposed strategies, which involve deactivating a subset of cells and/or beams, reduce the level of coverage diversity. 
Figure \ref{fig:coverage_div} shows the reduction in the number of cells each UE can potentially connect to after different optimization strategies discussed in this work. 
The global joint optimization leads to least overall link SNR and coverage diversity owing to most cell/beam deactivation.
Devising schemes to find an optimal balance between network energy savings and diversity is left for future work.

\subsubsection{Cell Search}
In idle mode, UEs need to periodically search for and measure neighboring cells and execute cell reselection when necessary to ensure they always have a strong and reliable candidate cell for upcoming connections. 
The cell search process is the main factor in UE energy consumption during idle mode. A positive consequence of cell/beam deactivation is the reduction in the number of candidate cells/beams that idle UEs need to search for. 
Figure~\ref{fig:beam_per_sector} shows a distribution of the number of active beams per cell, which is reduced from the baseline of 32 beams to at most 9 beams. This translates to up to 3X reduction in the UEs' search and measurement window, providing an opportunity for UE energy savings. 
\vspace{-2pt}




\section{Conclusion}
\label{sec:conclusion}
\vspace{-2pt}
This work presents several network energy-saving strategies by optimizing network resources at local and global levels during low-traffic or idle-mode operations in mm-Wave band. These solutions advance the literature on energy-efficient operation in 5G and beyond. Using the developed digital twin, our analysis demonstrates significant improvements in the energy and cost efficiency, as well as the sustainability, of telecommunication networks.
In the future, the proposed optimization and digital twin framework can be expanded to incorporate different objectives and support multiple levels of traffic load and multiple frequency layers.
\section{Acknowledgement}
The authors would like to thank Junyi Li (Engineering VP and Qualcomm Fellow), Tao Luo (Engineering VP), and Cihat Kececi (Senior Engineer) of Qualcomm Technologies Inc. for their insights, feedback, and support.




\bibliographystyle{IEEEtran}
\bibliography{references}

\end{document}